\documentclass[lettersize,journal,final]{IEEEtran}
\usepackage{amsmath,amssymb,amsfonts}
\usepackage{cuted}
\usepackage{algorithmic}
\usepackage{algorithm}
\usepackage{array}
\usepackage[caption=true,font=normalsize,labelfont=sf,textfont=sf]{subfig}
\usepackage{textcomp}
\usepackage{stfloats}
\usepackage{url}
\usepackage{verbatim}
\usepackage{graphicx}
\usepackage{cite}
\usepackage[table]{xcolor}
\usepackage{import}
\usepackage{steinmetz}
\usepackage{placeins}
\usepackage{setspace}
\usepackage{framed}
\usepackage{multirow}

\IEEEoverridecommandlockouts
\graphicspath{{figures}}
\definecolor{nb}{rgb}{0.2,0.2,0.702}
\input{tcilatex}
\begin{document}

\title{Multitarget Bistatic MIMO RADAR}
\author{Nadeem Dar and Athanassios Manikas \\
Department of Electrical and Electronic Engineering\\
Imperial College London }
\maketitle

\begin{abstract}
This paper is concerned with the investigation of the bistatic MIMO radar
for estimating various multitarget parameters of interest in the presence of
clutter and noise. The parameters of interest include Direction of Departure
(DOD), Direction of Arrival (DOA), range and velocity and a novel algorithm
is proposed for estimating these target parameters based on the concepts of
the "array manifold" and "manifold extenders". The performance of the
proposed algorithm is evaluated using Monte Carlo simulation studies.
\end{abstract}

\markboth{Journal of \LaTeX\ Class Files,~Vol.~14, No.~8, May~2024}
{Shell \MakeLowercase{\textit{et al.}}: A Sample Article Using IEEEtran.cls for IEEE Journals}
\IEEEpubid{\begin{minipage}{\textwidth}\ \\[12pt]
978-1-4799-5500-8/14/\$31.00 \copyright 2024 IEEE
\end{minipage}}%

\begin{IEEEkeywords}
Bistatic MIMO Radar, Direction of Arrival, Direction of Departure, Array Manifold, Manifold Extenders%
\end{IEEEkeywords}

\begin{tabular}{ll}
& 
\hspace{2em}\textsc{Notations}
\\ 
$a,$ $A$ & Scalar \\ 
$\underline{a},$ $\underline{A}$ & Column Vector \\ 
$\mathbf{a},$ $\mathbf{A},\mathbb{A}$ & Matrix \\ 
$\left( \mathbb{\cdot }\right) ^{T},\left( \mathbb{\cdot }\right) ^{H}$ & 
Transpose, Hermitian transpose \\ 
$\left( \mathbb{\cdot }\right) ^{\#}$ & Pseudo-inverse \\ 
$\left( \mathbb{\cdot }\right) ^{\ast }$ & Conjugate \\ 
$\mathcal{E}\{\cdot \}$ & Expectation operator \\ 
$\otimes $ & Kronecker product \\ 
$\odot $ & Hadamard product \\ 
$\circledast $ & C$\text{onvolution}$ \\ 
$\forall $ & "for all" \\ 
$\underline{1}_{N}$ & Column vector of $N$ ones \\ 
$\underline{0}_{N}$ & Column vector of $N$ zeros \\ 
$\mathbb{O}_{M\times N}$ & Matrix of zeros of size $M\times N$ \\ 
$\mathbb{I}_{N}$ & $N\times N$ Identity matrix \\ 
$\mathcal{R},$ $\mathcal{C}$ & Set of real \& complex numbers \\ 
$\underline{\func{col}}_{\ell }$ & $\ell $-th column of a matrix%
\end{tabular}

\section{Introduction}

\IEEEpubidadjcol%

Modern radars operate in the presence of objects/targets, clutter, noise and
interferences \cite{Richards2010_v1}. The main objectives of a radar are to
detect the presence of targets and estimate their parameters, especially
when targets are located close together in space. Once the presence of a
target is detected in clutter, then the aims are:

\begin{itemize}
\item to accurately estimate various target parameters, such as range,
direction and velocity, and

\item to solve the targets' "classification" problem, i.e. identify the type
of target by getting its "electronics signature".
\end{itemize}

The simplest radar architecture is the \textquotedblleft
monostatic\textquotedblright\ \cite{Melvin2013a_v2}\ where the Tx and Rx
antennas are \textquotedblleft collocated\textquotedblright , and thus the
DOD (Direction of Departure) from the radar's Tx to the target is identical
to the DOA (Direction of Arrival) from the target to the radar's Rx. Some
popular examples of DOA estimation for monostatic MIMO radar are the Least
Square (LS), Capon and Amplitude and Phase Estimation (APES) MIMO algorithms 
\cite{Luo2013}.

This paper is concerned with "bistatic" radar where there is some
considerable distance between the radar's Tx and Rx. Consequently, the DOA
is different from the DOD. Bistatic radars introduce several technical
complications but there are many significant advantages relative to
monostatic radar architectures. For instance, bistatic radar can handle
stealthy targets while \textquotedblleft jamming\textquotedblright\ is very
difficult. The most powerful and modern monostatic/bistatic radar is the
MIMO radar which employs antenna arrays of known geometries at both the Tx
and Rx sides. As the title of the paper suggests, this work is concerned
with bistatic MIMO radar with emphasis given to DOA and DOD estimation.
Existing DOA-DOD estimation algorithms in bistatic MIMO radar have been
reported in \cite{LaidBencheikh2010,Wen2020a,Cui2018,Sakhnini2023}.

In \cite{LaidBencheikh2010}, a joint DOA and DOD estimation algorithm is
proposed for bistatic MIMO radar based on a polynomial root-finding subspace
method. The main drawback of \cite{LaidBencheikh2010} is that the proposed
algorithm is restricted to bistatic MIMO where both Tx and Rx arrays are
uniform linear arrays.

In \cite{Wen2020a}, a two-dimensional DOA and DOD estimation is proposed
based on parallel factor (PARAFAC) analysis instead of Eigenvector
decomposition. The paper claims the computational efficiency using tensor
processing, but no computational complexity comparison with any conventional
method is presented.

In \cite{Cui2018}, a DOA-DOD and Range estimation is presented for Frequency
Diverse Array FDA-MIMO radar. This is a specific form of MIMO radar which
suffers from the "ambiguity problem" where a single target appears to have
many ranges and many DOD. Thus, \cite{Cui2018} is mainly focused on
resolving the ambiguity problem while it also assumes ULA for both Tx and Rx
arrays.

In \cite{Sakhnini2023}, the time synchronization is proposed between the Tx
and Rx sites. First, it estimates the bistatic range through matched
filtering and non-linear least squares (NLS) optimisation. Then the DOA and
DOD are calculated using the known bistatic geometry i.e., the bistatic
baseline, Tx, and Rx array locations. Finally, the synchronization offset is
calculated through the estimated/calculated parameters variance. This
calculation is only possible in the presence of a single target, and the
system requires a high SNR to operate reasonably.

As stated before, in this paper DOA and DOD are the primary parameters to be
estimated based on the concept of "extended manifolds" which are functions
of the array manifold. For plane-wave propagation, the array manifold is a
vector function of the target direction, array geometry and carrier
frequency. Array manifolds (or simply manifolds) are mathematical objects
such as \textquotedblleft curves\textquotedblright , \textquotedblleft
surfaces\textquotedblright\ etc. (i.e., non-linear subspaces). These
manifolds have been analysed in \cite{Manikas2004} and their shapes and
parameters are directly connected to the accuracy, resolution, and detection
capabilities of the array.%
\begin{figure*}
\centering
\includegraphics[width=0.90\textwidth]{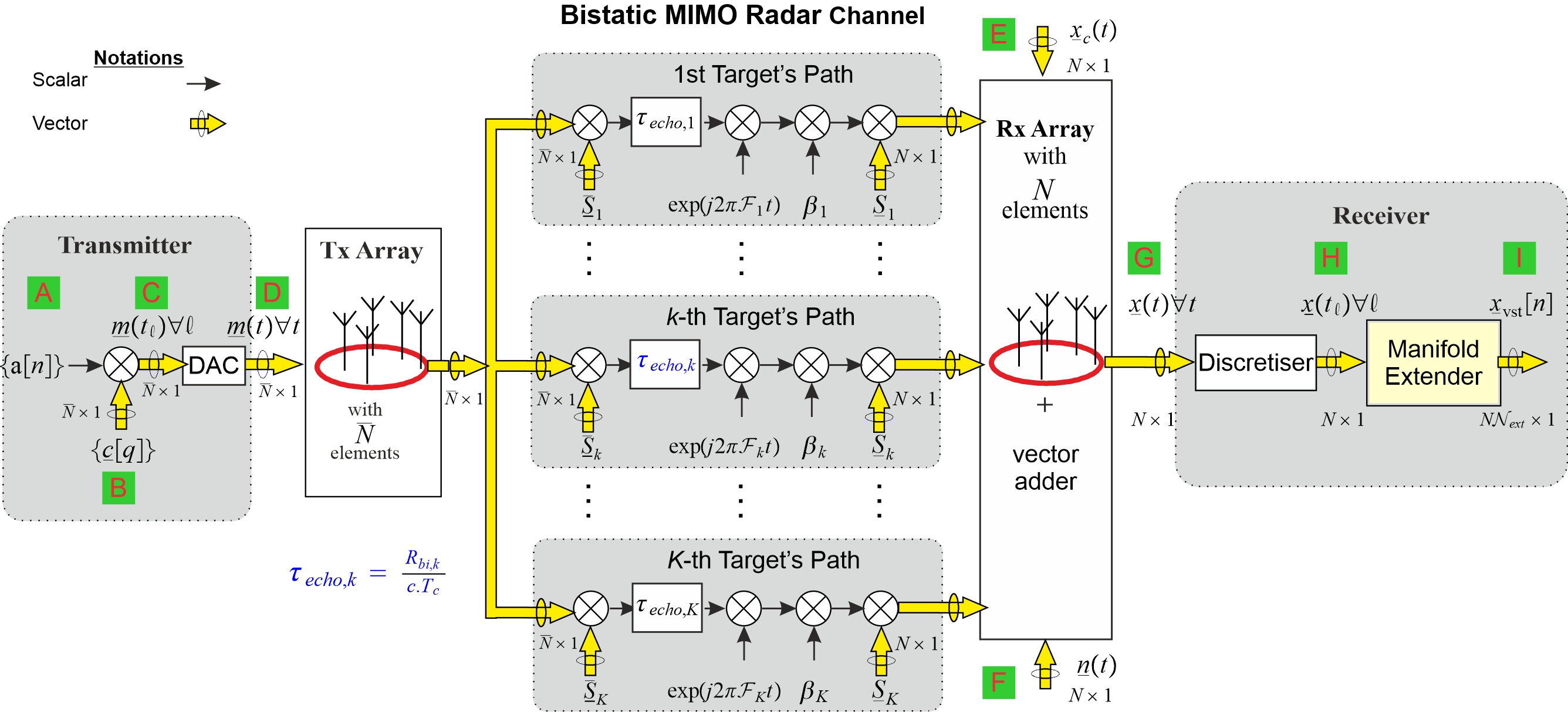}
\caption{Multi-Target MIMO Radar}
\label{fig_ch_model_01}
\end{figure*}%

These days more radar system and target parameters may be added to the array
manifold beyond the array geometry, the target direction and the carrier
frequency. These new manifolds with additional parameters of interest are
known as extended manifolds \cite{Efstathopoulos2011} to separate them from
the standard manifold which for convenience are also known as
\textquotedblleft spatial\textquotedblright\ array manifolds. Such
additional parameters may be polarisation, Doppler, delay, PN-codes, and
multi-carriers, etc. Every time a new system parameter is added, instead of
having a complex analysis like the one presented in \cite{Manikas2004}, the
aim is to express the extended manifold vector as a function of the spatial
manifold vector and make a direct connection to the properties and
capabilities between the new manifold and the original spatial manifold.

In this paper, the "manifold extender" will be used by starting with the Rx
spatial array manifold and then forming the Rx's extended manifold by adding
the following three new but unknown parameters:

\begin{itemize}
\item the target's range-bin,

\item Doppler frequency (for moving targets),

\item The DOD from the Tx to the target.
\end{itemize}

This paper builds on the DOA and DOD estimation proposed in \cite{Tang2022}
for a MIMO communication system. It explores the virtual array concept (a
form of extended array manifold \cite{Efstathopoulos2011, Commin2012}) to
extend the system's observation space from $N$ (number of Rx antennas) to $N%
\overline{N}$ (i.e. the product of Rx antennas $N$ and the Tx antennas $%
\overline{N}$ ), hence enabling to detect and process more targets than Rx
antenna elements with greater resolution capabilities\footnote{%
The concept of the extended array manifold has been also used in \cite%
{Ren2019} for monostatic MIMO radar and showed that the extended array
manifold algorithms outperformed the conventional algorithms/processing with
respect to RMSE criterion.}.

In summary, this paper builds upon the ideas put forth in previous works,
including those by \cite{Efstathopoulos2011, Commin2012, Ren2019, Tang2022},
and extends the concept to Multitarget Bistatic MIMO radar. The paper is
structured as follows: In Section 2, the generation of the Tx baseband
signal of the proposed system is presented. In Section 3, the MIMO
multitarget channel model that will be used in this paper is discussed.
Section 4 is concerned with the modelling of the received array vector
signal. Section 5 focuses on the proposed parameter estimation algorithms,
while Section 6 details the computer simulation environments and results.
The paper is concluded in Section 7.

\section{Tx Signal Modelling}

\IEEEPARstart{F}{igure}
\ref{fig_ch_model_01} conceptually illustrates a bistatic MIMO radar
architecture with the yellow arrow representing vectors of parallel signals.
At Point-A of the transmitter (Tx), a sequence of $N_{s}$ symbols $\{a[n]\in
\pm 1,\forall n=1,2,\ldots N_{s}\}$ is used with pulse duration $T_{p}$. The
Tx uses an antenna array of $\overline{N}$ antennas and the sequence $%
\{a[n]\in \pm 1\}$ is copied to all the $\overline{N}$ RF units. Then this
sequence is spread over a larger bandwidth by using a vector of $\overline{N}
$ PN-codes (one PN-code per Tx-antenna) as shown at Point-B. All these
PN-codes have length $\mathcal{N}_{c}$ and chip period $T_{c}$ with

\begin{equation}
\mathcal{N}_{c}T_{c}=T_{p}
\end{equation}%
The radar's $\mathrm{PRI}$ (Pulse Repetition Interval) is assumed having $%
N_{p}$ uncompressed range bins or, equivalently, $N_{p}\mathcal{N}_{c}$
compressed range bins. That is 
\begin{figure*}[tbp]
\centering
\includegraphics[width=0.95\textwidth]{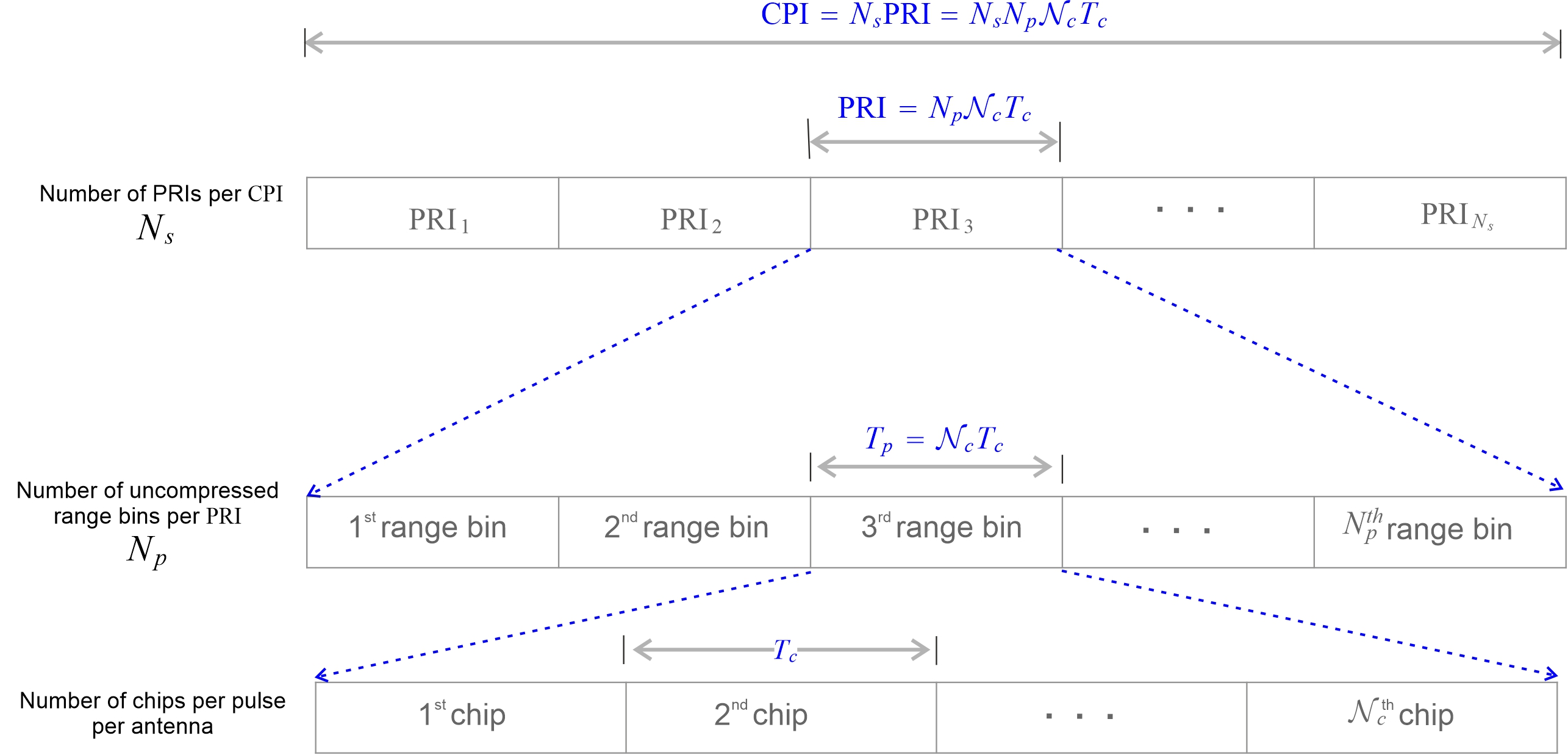}
\caption{Baseband Tx Signal Generation}
\label{fig_bb_sig_01}
\end{figure*}

\begin{eqnarray}
\mathrm{PRI} &=&N_{p}T_{p} \\
&=&N_{p}\mathcal{N}_{c}T_{c}
\end{eqnarray}

Figure \ref{fig_bb_sig_01} illustrates the above time frame structure of the
radar, considering also that the CPI (coherent processing interval) over
which the radar will estimate the various target parameters has a duration
equal to $N_{s}\mathrm{PRI}$. That is: 
\begin{eqnarray}
\mathrm{CPI} &=&N_{s}\mathrm{PRI}=N_{s}N_{p}\mathcal{N}_{c}T_{c} \\
&=&N_{s}N_{p}\mathcal{N}_{c}T_{c}
\end{eqnarray}%
The number of data symbols generated at Point-C of Figure \ref%
{fig_ch_model_01} over one CPI can be modelled by the following matrix $%
\mathbb{M}$: 
\begin{eqnarray}
\mathbb{M} &\triangleq &\left[ 
\begin{tabular}{llll}
$\underline{m}_{1},$ & $\underline{m}_{2},$ & $\ldots ,$ & $\underline{m}%
_{N_{s}N_{p}\mathcal{N}_{c}}$%
\end{tabular}%
\right] \\
&=&\underline{\mathrm{a}}^{T}\otimes \underset{\triangleq \mathbb{C}%
_{ex}^{T}\in \mathcal{R}^{\overline{N}\times N_{p}\mathcal{N}_{c}}}{%
\underbrace{%
\begin{bmatrix}
\mathbb{C}, & \mathbb{O}_{\overline{N}\times (N_{p}\mathcal{N}_{c}-\mathcal{N%
}_{c})}%
\end{bmatrix}%
}}\in \mathcal{R}^{\overline{N}\times N_{s}N_{p}\mathcal{N}_{c}}
\label{eq_M_b}
\end{eqnarray}%
where the data symbols vector $\underline{\mathrm{a}}$ is defined as 
\begin{equation}
\underline{\mathrm{a}}\triangleq 
\begin{bmatrix}
\mathrm{a}[1], & \mathrm{a}[2], & \ldots , & \mathrm{a}[N_{s}]%
\end{bmatrix}%
^{T}\text{{\small \ }}\in \mathcal{R}^{N_{s}\times 1}
\end{equation}%
and the PN-codes matrix has been extended with zeros until the end of each
PRI, i.e. $N_{p}\mathcal{N}_{c}-\mathcal{N}_{c}$ columns of $\overline{N}$
zeros

\scalebox{0.8}{
	\begin{minipage}{0.90\columnwidth}
		\begin{eqnarray}
			\mathbb{C}_{ex}^{T} &=&\begin{bmatrix}
				\underline{c}[1], & \underline{c}[2], & \ldots , & \underline{c}[q], & 
				\ldots , & \underline{c}[N_{p}\mathcal{N}_{c}]\end{bmatrix}\in \mathcal{R}^{\overline{N}\times N_{p}\mathcal{N}_{c}}  \notag \\
			&\mathbb{=}&\underset{\triangleq \text{\scalebox{1.2}{$\mathbb{C}$}}^{T}}{\left[ \underbrace{\begin{array}{cccc}
						\alpha _{1}[1], & \alpha _{1}[2], & \cdots  & \alpha _{1}[\mathcal{N}_{c}],
						\\ 
						\alpha _{2}[1], & \alpha _{2}[2], & \cdots  & \alpha _{2}[\mathcal{N}_{c}],
						\\ 
						\vdots  & \vdots  &  & \vdots  \\ 
						\alpha _{_{\overline{N}}}[1], & \alpha _{_{\overline{N}}}[2], & \cdots  & 
						\alpha _{\overline{N}}[\mathcal{N}_{c}],\end{array}}\right. }\underset{\begin{array}{c}
						\text{ \scalebox{0.8}{$N_{p}\mathcal{N}_{c}-\mathcal{N}_{c}$}} \\
						\text{ \scalebox{0.8}{columns of $\overline{N}$ zeros}}\end{array}}{\left. \underbrace{\begin{array}{cccc}
						0, & 0, & \cdots  & 0 \\ 
						0, & 0, & \cdots  & 0 \\ 
						\vdots  & \vdots  &  & \vdots  \\ 
						0, & 0, & \cdots  & 0\end{array}}\right]}   \notag
	\end{eqnarray}\end{minipage}
}%
\begin{equation}
\end{equation}

As shown at Point-B of Figure \ref{fig_ch_model_01}, the $q$-th column of
the $\mathbb{C}_{ex}^{T}$\ above matrix is the vector $\underline{c}[q]$.
The $\ell $-th column\footnote{%
Note that $\ell $ is an integer taking values $1,2,...,N_{s}N_{p}\mathcal{N}%
_{c}.$} of the matrix $\mathbb{M}$ (i.e. $m_{\ell }$) representing the
symbols that appear at Point-C during the chip interval $\ell T_{c}$. It is
important to point out that the PN-code matrix $\mathbb{C}$ has a covariance
matrix approximately equals to an identity matrix (almost orthogonal
PN-codes). That is:

\begin{equation}
\mathbb{R}_{\mathbb{CC}}=\frac{1}{\mathcal{N}_{c}}\mathbb{C}^{T}\mathbb{C}%
\approx \mathbb{I}\overline{_{N}}\in \mathcal{R}^{\overline{N}\times 
\overline{N}}
\end{equation}%
Then it is easy to see (using Equation \ref{eq_M_b}) that the covariance
matrix for the data matrix $M$ is%
\begin{equation}
\mathbb{R}_{\mathbb{MM}}=\frac{1}{\mathcal{N}_{c}.N_{s}}\mathbb{MM}%
^{T}\approx \mathbb{I}\overline{_{N}}\in \mathcal{R}^{\overline{N}\times 
\overline{N}}
\end{equation}%
At Point-D in Figure \ref{fig_ch_model_01}, after the DAC (digital to
analogue converter) block, the analogue baseband transmitted vector signal
is represented as follows:%
\begin{equation}
\underline{m}(t)=%
\begin{bmatrix}
m_{1}(t), & m_{2}(t), & ..., & m_{\overline{N}}(t)%
\end{bmatrix}%
^{T}  \label{eq_tx_mt}
\end{equation}

\section{Radar Channel Model}

\begin{figure}[]
\centering
\includegraphics[width=0.90\columnwidth]
{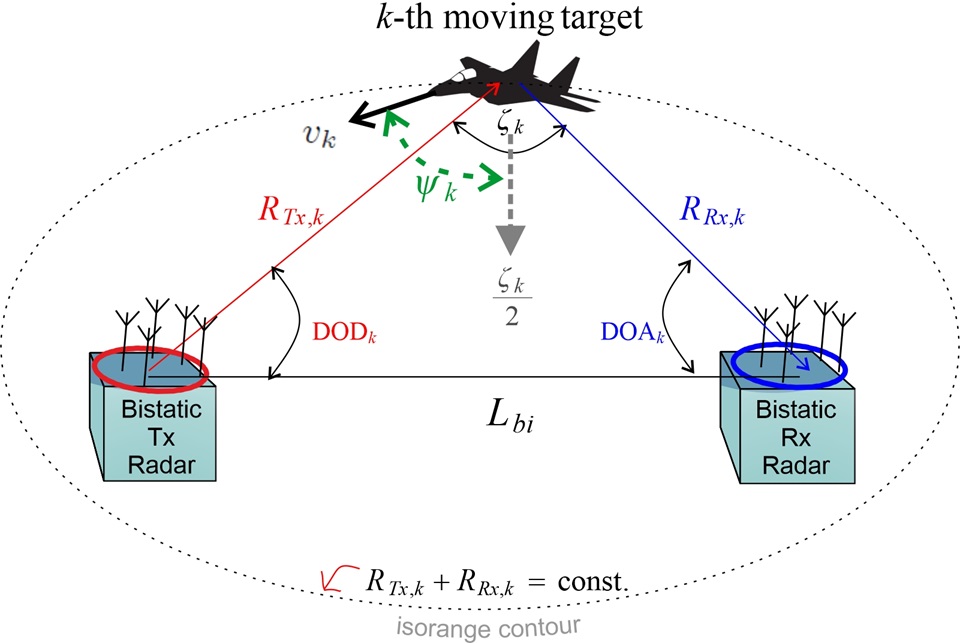}
\caption{Bistatic Radar's Target}
\label{key_fig_bi_velocity}
\end{figure}

The wireless MIMO channel is modelled as shown in Figure \ref%
{fig_ch_model_01} between Points-D and G. It considers a bistatic MIMO pulse
radar operating in the presence of $K$ moving targets. The geometry of the $%
k $-th moving target is shown in Figure \ref{key_fig_bi_velocity} having
velocity $v_{k}$ at a bistatic range $R_{bi,k}$. The vector-signal at the
output Point-D in Figure \ref{fig_ch_model_01} is transformed by the Tx
antenna array to a vector of $\overline{N}$ electromagnetic waves (one per
antenna) which will propagate through our 3D physical space. The various
"displacements" of these electromagnetic waves at the Tx antennas are a
function of

\begin{itemize}
\item the Tx-array geometry, and

\item their direction of propagation $\left( \overline{\theta }_{k},%
\overline{\phi }_{k}\right) $, where $\overline{\theta }_{k}$ represents
azimuth and $\overline{\phi }_{k}$ elevation angles.
\end{itemize}

\noindent These electromagnetic "displacements" for the $k$-th target's
direction $(\overline{\theta }_{k},\overline{\phi }_{k})$ are modelled by
the Tx manifold vector: 
\begin{eqnarray}
\underline{\overline{S}}_{k} &\triangleq &\underline{\overline{S}}(\overline{%
\theta }_{k},\overline{\phi }_{k})\in \mathcal{C}^{\overline{N}\times 1} 
\notag \\
&=&\exp \left( +j%
\begin{bmatrix}
\overline{\underline{r}}_{x}, & \overline{\underline{r}}_{y}, & \overline{%
\underline{r}}_{z}%
\end{bmatrix}%
\underline{k}(\overline{\theta }_{k},\overline{\phi }_{k})\right)
\end{eqnarray}%
where 
\begin{eqnarray}
\begin{bmatrix}
\overline{\underline{r}}_{x}, & \overline{\underline{r}}_{y}, & \overline{%
\underline{r}}_{z}%
\end{bmatrix}
&=&\text{Tx antenna array geometry}  \notag \\
&&  \label{eq_tx_r_xyz} \\
\underline{k}(\overline{\theta }_{k},\overline{\phi }_{k}) &=&\frac{2\pi }{%
\lambda }\underline{u}(\overline{\theta }_{k},\overline{\phi }_{k})\text{ is
wavenumber vector}  \notag \\
&&  \label{eq_tx_k}
\end{eqnarray}%
with $\underline{u}(\overline{\theta }_{k},\overline{\phi }_{k})$ being a
unity norm vector pointing towards the direction of departure, i.e.

\begin{equation}
\underline{u}(\overline{\theta }_{k},\overline{\phi }_{k})=%
\begin{bmatrix}
\cos \overline{\theta }_{k}\cos \overline{\phi }_{k} \\ 
\sin \overline{\theta }_{k}\cos \overline{\phi }_{k} \\ 
\sin \overline{\phi }_{k}%
\end{bmatrix}
\label{eq_tx_u}
\end{equation}

The transmitted electromagnetic waves will be reflected by the $k$-th target
and the echo will be received by the radar's Rx antenna array after a
propagation time 
\begin{eqnarray}
\tau _{echo,k} &=&\frac{R_{T_{x},k}+R_{R_{x},k}}{c},\forall k
\label{eq_techo} \\
d_{k} &=&\left\lfloor \frac{\tau _{echo,k}}{T_{c}}\right\rfloor
\end{eqnarray}%
where $c$ is the speed of light. Figure \ref{fig_ch_model_01} shows the
delay $\tau _{echo,k}$, the path gain $\beta _{k}$ and the effects of the
target's Doppler frequency $\mathcal{F}_{k}$. \ Note that the parameter $%
\beta _{k}$ in Figure \ref{fig_ch_model_01} is modelled as follow:%
\begin{equation}
\beta _{k}=\left\vert \beta _{k}\right\vert \exp (j\text{%
$\phase{\beta_k}$%
)}  \label{eq_Beta_01}
\end{equation}

\noindent where%
\begin{eqnarray}
\left\vert \beta _{k}\right\vert &=&\sqrt{\frac{G_{Tx}G_{Rx}}{(4\pi )^{3}}.}%
\left( \frac{\lambda }{R_{Tx}R_{Rx}}\right) .\sqrt{\mathrm{RCS}_{k}} \\
\text{%
$\phase{\beta_k}$%
} &=&-j2\pi F_{c}\frac{R_{bi,k}}{c}+j\psi _{k}+j2\pi \mathcal{F}_{k}t  \notag
\\
&&
\end{eqnarray}

\noindent $\psi _{k}$ is random phase and $F_{c}$\ is RF carrier frequency.
Thus the parameter $\beta _{k}$ of the $k$-th target is a function of:

\begin{itemize}
\item the target's radar cross section ($\mathrm{RCS}_{k}$),

\item the antenna gains of the Tx and Rx, i.e. $G_{T_{x}}$ and $G_{R_{x}}$
respectively,

\item the Tx to target range $R_{T_{x},k}$ and target to Rx range $R_{R_{x}}$%
.
\end{itemize}

The channel model also considers targets' movement and resulting Doppler
shift on the Tx carrier frequency. In Figure \ref{fig_ch_model_01} the
Doppler shift of the $k$-th moving target is represented by the term $\exp
(j2\pi \mathcal{F}_{k}t)$. For the bistatic geometry of Figure \ref%
{key_fig_bi_velocity}, this Doppler shift is given as follows \cite%
{Richards2010_v1}:

{\normalsize 
\begin{equation}
\mathcal{F}_{k}=\frac{2v_{k}}{\lambda }\cos \psi _{bi,k}\cos \frac{\zeta _{k}%
}{2}
\end{equation}%
}

\noindent where the subscript $k$ signifies the $k$-th target, $v_{k}$
denotes its velocity, $\zeta _{k}$ is the angle between incident and
reflected wave, and $\psi _{bi,k}$ represents the angle between the
discontinuous $\frac{\zeta _{k}}{2}$ and the direction of movement.

In a similar fashion, for the $k$-th target, the various "displacements" of
the electromagnetic waves arriving at the Rx antenna array with the direction of
arrival $\theta _{k},\phi _{k}$ is modelled by the Rx manifold vector:%
\begin{eqnarray}
\underline{S}_{k} &\triangleq &\underline{S}(\theta _{k},\phi _{k})\in 
\mathcal{C}^{N\times 1} \\
&=&\exp \left( -j%
\begin{bmatrix}
\underline{r}_{x}, & \underline{r}_{y}, & \underline{r}_{z}%
\end{bmatrix}%
\underline{k}(\theta _{k},\phi _{k})\right)
\end{eqnarray}%
where the matrix $\left[ \underline{r}_{x},\underline{r}_{y},\underline{r}%
_{z}\right] ^{T}$ denotes the Cartesian coordinates of the Rx antenna array
elements and $\underline{k}(\theta _{k},\phi _{k})$\ is the wavelength
vector corresponding to the Direction of Arrival (DOA).

Based on the above, the impulse response (IR) of a MIMO radar channel can be
summarised using the following equation:

{\normalsize 
\begin{equation}
\text{IR}(t)=\overset{K}{\sum_{k=1}}\beta _{k}\exp (j2\pi \mathcal{F}_{k}t)%
\underline{S}_{k}\overline{\underline{S}}_{k}^{H}\underline{\delta }(t-\tau
_{echo,k})
\end{equation}%
}

\section{Radar Receiver based on "Manifold Extender"}

At Point-G of Figure \ref{fig_ch_model_01}, based on the radar's channel
modelling presented in Section III, the analogue received vector-signal $%
\underline{x}(t)\in \mathcal{C}^{N\times 1}$ at the output of the Rx antenna
array can be written as: 
\begin{eqnarray}
\underline{x}(t) &=&%
\begin{bmatrix}
x_{1}(t), & x_{2}(t), & ..., & x_{N}(t)%
\end{bmatrix}%
^{T}\in \mathcal{C}^{N\times 1}  \label{eq_xt_01} \\
&=&\overset{K}{\sum_{k=1}}\underset{\text{scalar signal (echo) from the }k%
\text{-th target}}{\underbrace{\sqrt{P_{Tx}}\beta _{k}\exp (j2\pi \mathcal{F}%
_{k}t)\overline{\underline{S}}_{k}^{H}\underline{m}(t-\tau _{echo,k})}}%
\underline{S}_{k}  \notag \\
&&+\underline{x}_{c}(t)+\underline{\mathrm{n}}(t)  \label{eq_xt_02}
\end{eqnarray}%
\noindent

\noindent where $\underline{x}_{c}(t)$ represents the "clutter" effects and $%
\underline{\mathrm{n}}(t)$ is the complex additive white Gaussian noise
(AWGN) of $\mathcal{CN}(\underline{0},\ \sigma _{\mathrm{n}}^{2}I_{N}),$
i.e. of zero mean and covariance matrix $\mathbb{R}_{\mathrm{nn}}=\sigma _{%
\mathrm{n}}^{2}I_{N}$. Remember that the subscript $k$ refers to the $k$-th
target and a bar on top of a symbol represents a Tx parameter.

The analogue vector-signal $\underline{x}(t)$ at Point-G of Figure \ref%
{fig_ch_model_01} is then digitised to produce at Point-H the $\ell $-th
snapshot vector $\underline{x}(t_{\ell }),\forall \ell =1,2,\ldots $ , which
is represented as:%
\begin{equation}
\underline{x}(t_{\ell })=%
\begin{bmatrix}
x_{1}(t_{\ell }), & x_{2}(t_{\ell }), & ..., & x_{N}(t_{\ell })%
\end{bmatrix}%
^{T}\in \mathcal{C}^{N\times 1}
\end{equation}

These snapshots, $\forall \ell ,$ at Point-H are then driven to a suitable
"manifold extender" block which is going to be employed by the proposed
novel algorithm for estimating both DOA and DOD of all targets. 
\begin{figure*}
\centering
\includegraphics[width=0.95\textwidth]{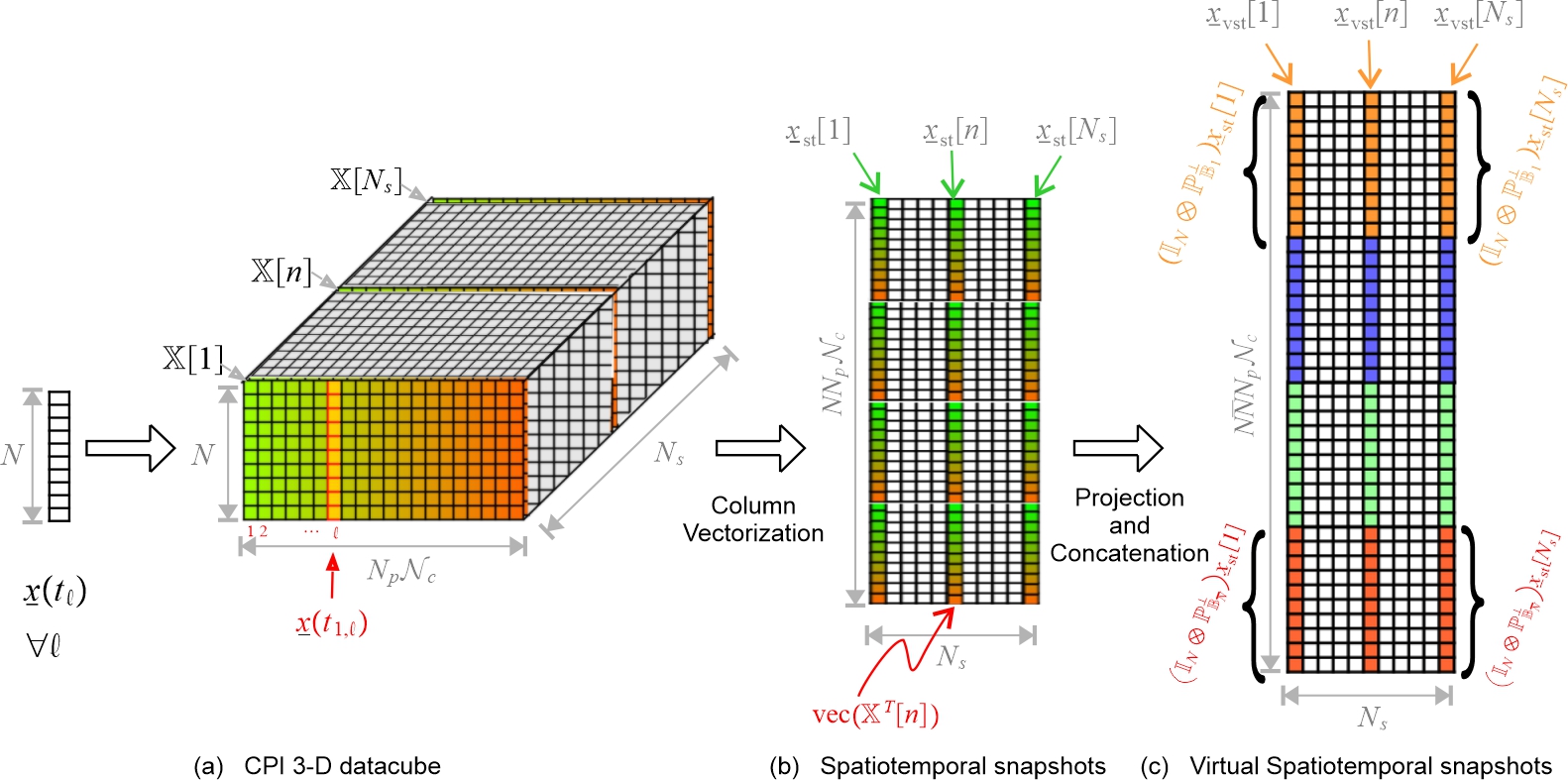}
\caption{MIMO Radar Rx 3-D Data Cube $\QTR{Bbb}{X}$ for One CPI and Extended Snapshots Formation}
\label{fig_3d_cube_X}
\end{figure*}%

\subsection{Manifold Extender}

The manifold extender involves a complex mapping of spatial manifold to
extended manifold using additional system parameters. In this paper these
additional parameters are the DOD, radar's range bins, the PN-code \ matrix $%
\mathbb{C}$ and the Doppler frequency. That is 
\begin{gather}
\begin{tabular}{|l|}
\hline
$\underline{S}_{k}\text{: {\small {is a function of }}\textrm{DOA}}_{k}\text{%
, }F_{c}\text{, }c\text{ \text{{\small {and array geometry}}}}\color{white}%
............$ \\ \hline
\end{tabular}
\label{eq_Sk_expla} \\
\begin{tabular}{|l|}
\hline
$\underline{h}_{k}\text{: {\small {is a function of }}}\underline{S}_{k}%
\text{ {\small (see Equ. \ref{eq_Sk_expla}) plus}\ \textrm{DOD}}_{k}\text{, }%
\mathbb{C}\text{, }\ell _{k}\text{, }\mathcal{F}_{k}$ \\ \hline
\end{tabular}
\label{eq_hk_expla_}
\end{gather}

\noindent This implies that:

\begin{itemize}
\item at Point-H of Figure \ref{fig_ch_model_01}: $\underline{x}(t_{\ell
})\in \mathcal{C}^{N\times 1}$, with
\end{itemize}

\begin{equation}
\underline{x}(t_{\ell })=\left( \overset{K}{\sum_{k=1}}\text{function of }%
\underline{S}_{k}\right) +\underline{\mathrm{clutter}}+\text{ }\underline{%
\mathrm{noise}}  \label{eq_xtl_expla_01}
\end{equation}

\begin{itemize}
\item at Point-I of Figure \ref{fig_ch_model_01}: $\underline{x}_{\mathrm{vst%
}}[n]\in \mathcal{C}^{NN_{ext}\times 1}$, with{\normalsize 
\begin{eqnarray}
\underline{x}_{\mathrm{vst}}[n] &=&\left( \overset{K}{\sum_{k=1}}\text{%
function of }\underline{h}_{k}\right) +\underline{\mathrm{clutter}}+\text{ }%
\underline{\mathrm{noise}}\text{ }  \notag \\
&&  \label{eq_xvst_expla}
\end{eqnarray}%
}where%
\begin{equation*}
N_{ext}=\overline{N}N_{p}\mathcal{N}_{c}
\end{equation*}%
To go from Equation \ref{eq_xtl_expla_01} to Equation \ref{eq_xvst_expla}
the input data at Point-H should be re-arranged in a suitable way so that
Equation \ref{eq_Sk_expla} is transformed to Equation \ref{eq_hk_expla_}.
This re-arrangement of data is shown in Figure \ref{fig_3d_cube_X} and is
described as follows:

\item The snapshot at time instant $t_{\ell }$\ is denoted so far by $%
\underline{x}(t_{\ell })$\ where $\ell $ takes values $1,2,...,$ $N_{s}N_{p}%
\mathcal{N}_{c}$ representing the whole CPI. However, for every PRI we
collect its $N_{p}\mathcal{N}_{c}$ snapshot vectors. To reflect this
collection of snapshots to a specific PRI, the notation of the snapshot $%
\underline{x}(t_{\ell })$ with $\ell $ taking values $1,2,...,$ $N_{p}%
\mathcal{N}_{c}$ is updated by adding a second subscript, $n$, i.e. $%
\underline{x}(t_{n,\ell }),$ to indicate the $\ell ^{\text{th}}$ snapshot in
the $n^{\text{th}}$ PRI, with $\ell $ now taking values $1,2,\ldots ,$ $N_{p}%
\mathcal{N}_{c}$ and $n=1,2,\ldots ,$ $N_{s}$. This snapshot can be modelled
as:

\scalebox{0.8}{
\begin{minipage}{0.90\columnwidth}
		\begin{eqnarray}
			\underline{x}(t_{n,\ell }) &=&\overset{K}{\sum_{k=1}}\underset{\mathrm{scalar				}_{k_{n,\ell }}}{\underbrace{\sqrt{P_{Tx}}\beta _{k}\mathrm{a}[n]\exp (j2\pi 
					\mathcal{F}_{k}\ell T_{c})\underline{\overline{S}}_{k}^{H}\underline{\func{							col}}_{\ell }\left[ \left( \mathbb{J}^{d_{k}}\mathbb{C}_{ex}\right) ^{T}					\right] }}\underline{S}_{k}  \notag \\
			&&+\underline{x}_{c}(t_{n,\ell })+\underline{\mathrm{n}}(t_{n,\ell })  \notag
			\\
			&=&\overset{K}{\sum_{k=1}}\mathrm{scalar}_{k_{n,\ell }}.\underline{S}_{k}+			\underline{x}_{c}(t_{n,\ell })+\underline{\mathrm{n}}(t_{n,\ell })
			\notag
		\end{eqnarray}
\end{minipage}
}%
\begin{equation}  \label{eq_xtnl_01}
\end{equation}

where $\underline{\func{col}}_{\ell }\left[ \left( \mathbb{J}^{d_{k}}\mathbb{%
C}_{ex}\right) ^{T}\right] $ denotes the $\ell $-th column of the matrix $%
\left( \mathbb{J}^{d_{k}}\mathbb{C}_{ex}\right) ^{T}$ $\in \mathcal{R}^{%
\overline{N}\times N_{p}\mathcal{N}_{c}}$

\scalebox{0.8}{
\begin{minipage}{0.90\columnwidth}
\begin{equation*}
\left( \mathbb{J}^{d_{k}}\mathbb{C}_{ex}\right) ^{T}\mathbb{=}\underset{\begin{array}{c}
\text{\scalebox{0.8}{$d_k$ columns}}\\
\text{\scalebox{0.8}{$\overline{N}$ zeros}}\end{array}}{\left[ \underbrace{			\begin{array}{ccc}
				0, & \cdots  & 0, \\ 
				0, & \cdots  & 0, \\ 
				\vdots  &  & \vdots  \\ 
				0, & \cdots  & 0,			\end{array}		}\right. }\overset{\triangleq \text{\scalebox{1.2}{$\mathbb{C}$}}^{T}}{		\overbrace{			\begin{array}{ccc}
				\alpha _{1}[1], &  & \alpha _{1}[\mathcal{N}_{c}], \\ 
				\alpha _{2}[1], & \cdots  & \alpha _{2}[\mathcal{N}_{c}], \\ 
				\vdots  &  & \vdots  \\ 
				\alpha _{_{\overline{N}}}[1], & \cdots  & \alpha _{\overline{N}}[\mathcal{N}				_{c}],			\end{array}	}}\underset{		\begin{array}{c}
			\text{\scalebox{0.75}{$N_p\mathcal{N}_c-\mathcal{N}_c-d_k$}}\\
			\text{\scalebox{0.75}{columns of $\overline{N}$ zeros}}		\end{array}	}{\left. \underbrace{			\begin{array}{ccc}
				0, & \cdots  & 0 \\ 
				0, & \cdots  & 0 \\ 
				\vdots  &  & \vdots  \\ 
				0, & \cdots  & 0			\end{array}		}\right] }
\end{equation*}\end{minipage}
}%
\begin{equation}
\end{equation}
Note that the shifting matrix $\mathbb{J}\in \mathcal{R}^{N_{p}\mathcal{N}%
_{c}\times N_{p}\mathcal{N}_{c}}$ is defined \cite{Manikas2003} as%
\begin{equation*}
\mathbb{J}=%
\begin{bmatrix}
\underline{0}_{N_{p}\mathcal{N}_{c}-1}^{T}, & 0\ \ \ \ \  \\ 
\mathbb{I}_{N_{p}\mathcal{N}_{c}-1}, & \ \ \ \ \ \underline{0}_{N_{p}%
\mathcal{N}_{c}-1}%
\end{bmatrix}%
\end{equation*}

and the operation of the $\mathbb{J}^{d_{k}}\mathbb{C}_{ex}$ is down
shifting the columns of the code matrix $\mathbb{C}_{ex}$ by $d_{k}$
elements which is the amount of discrete delay $d_{k}$ corresponding to the $%
k$-th target.

\item the snapshots are assembled first into PRIs and then CPI interval, as
shown in Figure \ref{fig_3d_cube_X}a. The $\ell ^{\text{th}}$\ snapshot of
the $1^{\text{st}}$ PRI is also shown in Figure \ref{fig_3d_cube_X}a as $%
\underline{x}(t_{1,\ell })$. Thus during an observation interval of $N_{p}%
\mathcal{N}_{c}$ snapshots, corresponding to data matrix of the $n^{\text{th}%
}$ PRI $\mathbb{X}[n]\in \mathcal{C}^{N\times N_{p}\mathcal{N}_{c}}$ can be
modelled as follows:

\scalebox{0.85}{
\begin{minipage}{0.85\columnwidth}
\begin{eqnarray}
	\mathbb{X}[n] &=&	\begin{bmatrix}
		\underline{x}(t_{n,1}), & ..., & \underline{x}(t_{n,\ell }), & ..., & 
		\underline{x}(t_{n,N_{p}\mathcal{N}_{c}})	\end{bmatrix}
	\notag \\
	&=&\overset{\text{(echos from }K\text{-targets)}}{\overbrace{\overset{K}{				\sum_{k=1}}\sqrt{P_{Tx}}\beta _{k}\mathrm{a}[n]\underline{S}_{k}\underline{				\overline{S}}_{k}^{H}\left( \overset{\mathbb{\triangleq T}(d_{k},\mathcal{F}				_{k})}{\overbrace{\left( \mathbb{J}^{d_{k}}\mathbb{C}_{ex}\right) ^{T}\odot
					\left( \underline{1}_{\overline{N}}\underline{\mathcal{F}}_{c,k}^{T}\right) }			}\right) }\text{ \ \ }}  \notag \\
	&&+\mathbb{X}_{c}[n]\text{ \ \ \ (clutter term)}  \notag \\
	&&+\mathbb{N}[n]\text{ \ \ \ (noise term)}  \notag
\end{eqnarray}\end{minipage}
} \vspace{-1.8em}%
\begin{equation}  \label{eq_rx_Xn_2}
\end{equation}%
\qquad\ \ 

where%
\begin{equation}
\underline{\mathcal{F}}_{c,k}\mathbb{\triangleq }\exp \left( j2\pi \mathcal{F%
}_{k}%
\begin{bmatrix}
1 \\ 
2 \\ 
\vdots \\ 
N_{p}\mathcal{N}_{c}%
\end{bmatrix}%
T_{c}\right) \in \mathcal{C}^{N_{p}\mathcal{N}_{c}\times 1}
\end{equation}

\scalebox{0.85}{
\begin{minipage}{0.85\columnwidth}
\begin{eqnarray*}
\mathbb{X}_{c}[n] &=&	\begin{bmatrix}
\underline{x}_{c}(t_{n,1}), & ..., & \underline{x}_{c}(t_{n,\ell }), & ...,& \underline{x}_{c}(t_{n,N_{p}\mathcal{N}_{c}})	\end{bmatrix}
\\
\mathbb{N}[n] &=&		\begin{bmatrix}
\underline{\mathrm{n}}(t_{n,1}), & ..., & \underline{\mathrm{n}}(t_{n,\ell}), & ..., & \underline{\mathrm{n}}(t_{n,N_{p}\mathcal{N}_{c}})\notag	\end{bmatrix}			
\end{eqnarray*}
\end{minipage}
} \newline

\item Vectorisation of Equation \ref{eq_rx_Xn_2} will provide the $n^{\text{%
th}}$ column (associated with $n^{\text{th}}$ PRI) of Figure \ref%
{fig_3d_cube_X}b represented by the vector \ $\underline{x}_{\mathrm{st}}[n]$
which is modelled as follows:%
\begin{eqnarray}
\underline{x}_{\mathrm{st}}[n] &=&\limfunc{vec}\left( \mathbb{X}%
^{T}[n]\right) \in \mathcal{C}^{NN_{p}\mathcal{N}_{c}\times 1}
\label{eq_xst_01} \\
&=&\text{\scalebox{0.85}{$\sum\limits_{k=1}^{K}\sqrt{P_{Tx}}\beta
_{k}\mathrm{a}[n]\underset{\text{
scalar}}{\underbrace{\underline{\overline{S}}_{k}^{H}\underline{1}_{
\overline{N}}}}{\left( \underset{\underline{h}_{\mathrm{st,}k}}{\underbrace{
\underline{S}_{k}\otimes \left( \mathbb{J}^{d_{k}}\underline{c}_{s}\odot
\underline{\mathcal{F}}_{c,k}\right) }}\right) } $}}  \notag
\end{eqnarray}

with the\ composite code vector $\underline{c}_{s}$\ defined as the addition
of all the $\overline{N}$ PN-codes that is: 
\begin{equation}
\underline{c}_{s}\mathbb{\triangleq C}_{ex}\underline{1}_{\overline{N}}\in 
\mathcal{R}^{N_{p}\mathcal{N}_{c}\times 1}
\end{equation}

\item Finally the $\limfunc{vec}\left( \mathbb{X}^{T}[n]\right) $ is
projected as follows to provide one column of Figure \ref{fig_3d_cube_X}c 
\begin{subequations}
\begin{eqnarray}
\underline{x}_{\mathrm{vst}}[n] &=&%
\begin{bmatrix}
\left( \mathbb{I}_{N}\otimes \mathbb{P}_{\mathbb{B}_{1}}^{\perp }\right) \\ 
\left( \mathbb{I}_{N}\otimes \mathbb{P}_{\mathbb{B}_{2}}^{\perp }\right) \\ 
\vdots \\ 
\left( \mathbb{I}_{N}\otimes \mathbb{P}_{\mathbb{B}_{m}}^{\perp }\right) \\ 
\vdots \\ 
\left( \mathbb{I}_{N}\otimes \mathbb{P}_{\mathbb{B}_{\overline{N}}}^{\perp
}\right)%
\end{bmatrix}%
\underset{\mathbb{\triangleq }\underline{x}_{\mathrm{st}}[n]}{\underbrace{%
\limfunc{vec}\left( \mathbb{X}^{T}[n]\right) }}  \label{eq_xvstn_01} \\
&=&\sum\limits_{k=1}^{K}\sqrt{P_{Tx}}\beta _{k}\mathrm{a}[n]\mathbb{P}_{%
\mathbb{B}}^{\perp }\underline{h}_{k}+\underline{x}_{\mathrm{v,c}}[n]+%
\underline{\mathrm{n}}_{\mathrm{v}}[n]  \notag \\
&&  \label{eq_xvstn_02}
\end{eqnarray}%
where 
\end{subequations}
\begin{equation}
\underline{h}_{k}\ \triangleq \underline{S}_{k}\otimes \underline{\overline{S%
}}_{k}^{\ast }\otimes \left( \mathbb{J}^{d_{k}}\underline{c}_{s}\odot 
\underline{\mathcal{F}}_{c,k}\right)
\end{equation}%
and $\mathbb{P}_{\mathbb{B}_{m}}^{\perp }\in \mathcal{C}^{N_{p}\mathcal{N}%
_{c}\times N_{p}\mathcal{N}_{c}\text{ }}$ is a complement subspace
projection operator to isolate the signal from $m$-th Tx antenna from the
rest and is defined as 
\begin{equation}
\mathbb{P}_{\mathbb{B}_{m}}^{\perp }\triangleq \mathbb{I}_{N_{p}\mathcal{N}%
_{c}}-\mathbb{B}_{m}\left( \mathbb{B}_{m}^{H}\mathbb{B}_{m}\right) ^{-1}%
\mathbb{B}_{m}^{H}  \label{eq_pbm_01}
\end{equation}%
with 
\begin{equation}
\mathbb{B}_{m}=%
\begin{bmatrix}
\mathbb{J}^{d_{1}}\mathbb{C}_{m}\odot \left( \underline{\mathcal{F}}_{c,1}%
\underline{1}_{\overline{N}-1}^{T}\right) \\ 
\mathbb{J}^{d_{2}}\mathbb{C}_{m}\odot \left( \underline{\mathcal{F}}_{c,2}%
\underline{1}_{\overline{N}-1}^{T}\right) \\ 
\vdots \\ 
\mathbb{J}^{d_{K}}\mathbb{C}_{m}\odot \left( \underline{\mathcal{F}}_{c,K}%
\underline{1}_{\overline{N}-1}^{T}\right)%
\end{bmatrix}%
^{T}  \label{eq_Bm_01}
\end{equation}%
note that the matrix $\mathbb{B}_{m}$ $\in \mathcal{C}^{N_{p}\mathcal{N}%
_{c}\times K\left( \overline{N}-1\right) \text{ }}$represents the part of
extended array response vector covering delay and Doppler parameters for $K$
targets. Where $\mathbb{C}_{m}$ $\in \mathcal{C}^{N_{p}\mathcal{N}_{c}\times
\left( \overline{N}-1\right) }$ is the code matrix $\mathbb{C}_{ex,m}$ with $%
m$-th antenna's code removed as shown below:%
\begin{equation}
\mathbb{C}_{m}=%
\begin{bmatrix}
\underline{c}_{1}, & \cdots , & \underline{c}_{m-1}, & \underline{c}_{m+1},
& \cdots , & \underline{c}_{\overline{N}}%
\end{bmatrix}%
\end{equation}%
A similar projection operator matrix $\mathbb{P}_{\mathbb{B}}^{\perp }\in 
\mathcal{C}^{N\overline{N}N_{p}\mathcal{N}_{c}\times N\overline{N}N_{p}%
\mathcal{N}_{c}}$ can be formed for every $m$, that is for all antennas, and
the overall operator for the whole array can be represented as follows
\begin{equation}
\mathbb{P}_{\mathbb{B}}^{\perp }=%
\begin{bmatrix}
\mathbb{I}_{N}\otimes \mathbb{P}_{\mathbb{B}_{1}}^{\perp }, & \mathbb{O}%
_{NN_{p}\mathcal{N}_{c}}, & \cdots , & \mathbb{O}_{NN_{p}\mathcal{N}_{c}},
\\ 
\mathbb{O}_{NN_{p}\mathcal{N}_{c}}, & \mathbb{I}_{N}\otimes \mathbb{P}_{%
\mathbb{B}_{2}}^{\perp }, & \cdots , & \mathbb{O}_{NN_{p}\mathcal{N}_{c}},
\\ 
\vdots & \vdots & \ddots & \vdots \\ 
\mathbb{O}_{NN_{p}\mathcal{N}_{c}}, & \mathbb{O}_{NN_{p}\mathcal{N}_{c}}, & 
\cdots , & \mathbb{I}_{N}\otimes \mathbb{P}_{\mathbb{B}_{\overline{N}%
}}^{\perp }%
\end{bmatrix}
\label{eq_pb_all}
\end{equation}%
Equation \ref{eq_xvstn_02} reveals that $\underline{x}_{\mathrm{vst}}[n]$ is
a function of the extended array manifold vector\ $\underline{h}_{k}$\ $\in 
\mathcal{C}^{N\overline{N}N_{p}\mathcal{N}_{c}\times 1}$
\end{itemize}

\subsection{Joint Range and Doppler Estimation}

First, the range $=d_{k}T_{c}c$ and the Doppler $\mathcal{F}_{k}\forall k,$
can be estimated using the following cost function:

\begin{equation}
(\underline{d},\underline{\mathcal{F}})=\arg \ \underset{d,\mathcal{F}}{\max 
}\ \xi _{1}(d,\mathcal{F}),\forall d,\mathcal{F}  \label{eq_rd_01}
\end{equation}%
where

\begin{equation}
\xi _{1}(d,\mathcal{F})=\frac{\det \left( \mathbb{T}(d,\mathcal{F})^{H}%
\mathbb{T}(d,\mathcal{F})\right) }{\det \left( \mathbb{T}(d,\mathcal{F})^{H}%
\mathbb{P}_{n}\mathbb{T}(d,\mathcal{F})\right) },\forall d,\mathcal{F}
\label{eq_rd_02}
\end{equation}%
with

\begin{equation}
\mathbb{T}(d,\mathcal{F})=\mathbb{J}^{d}\mathbb{C}_{ex}\odot \left( 
\underline{1}_{\overline{N}}\underline{\mathcal{F}}_{c}\right)
\end{equation}%
$\mathbb{T}(d,\mathcal{F})$ is the transformation matrix defined in Equation %
\ref{eq_rx_Xn_2}. Furthermore, the matrix $\mathbb{P}_{n}$ in Equation \ref%
{eq_rd_02}, is the projection operator to the noise subspace, spanned by the
"noise" eigenvalues of the data covariance matrix%
\begin{equation}
\mathbb{R}_{\mathbb{XX}}=\frac{1}{N.N_{s}}\underset{n=1}{\overset{N_{s}}{%
\sum }}\mathbb{X}[n]\mathbb{X}[n]^{H}
\end{equation}

\subsection{Joint DOA and DOD Estimation}

Using the above-estimated delay and Doppler the complement projection
operator $\mathbb{P}_{\mathbb{B}}^{\perp }$ can be constructed to form the
virtual spatiotemporal snapshot vector $\underline{x}_{\mathrm{vst}}[n]$.
This enables to estimate the DOA and DOD using the cost function below:

\begin{equation}
(\underline{\theta },\underline{\overline{\theta }})=\arg \underset{\forall
\left( \theta ,\overline{\theta }\right) }{\max }\ \left. \xi _{2}(\theta ,%
\overline{\theta })\right\vert _{d=d_{k,},\mathcal{F=F}_{k}},\forall \theta ,%
\overline{\theta }  \label{cf_doa_dod_01}
\end{equation}%
and $\xi _{2}(\theta ,\overline{\theta })$ can be defined as below:%
\begin{equation}
\xi _{2}(\theta ,\overline{\theta })\triangleq \underset{k=1}{\overset{K}{%
\sum }}\left( \frac{\left\Vert \mathbb{P}_{\mathbb{B}}^{\perp }\underline{h}%
_{k}(\theta ,\overline{\theta })\right\Vert ^{2}}{\left( \mathbb{P}_{\mathbb{%
B}}^{\perp }\underline{h}_{k}(\theta ,\overline{\theta })\right) ^{H}\mathbb{%
P}_{n\mathrm{v}}\left( \mathbb{P}_{\mathbb{B}}^{\perp }\underline{h}%
_{k}(\theta ,\overline{\theta })\right) }\right) ,\forall \theta ,\forall 
\overline{\theta }  \label{cf_doa_dod_02}
\end{equation}%
where $\mathbb{P}_{n\mathrm{v}}$ is the projection operator to the noise
subspace spanned by the "noise" eigenvector of the covariance matrix of the
virtual-spatiotemporal snapshots given by Equation \ref{eq_xvstn_01}

{\normalsize 
\begin{equation}
R_{xx}=\frac{1}{N_{s}}\underset{n=1}{\overset{N_{s}}{\sum }}\underline{x}_{%
\mathrm{vst}}[n]\underline{x}_{\mathrm{vst}}[n]^{H}
\end{equation}%
}

\section{Computer Simulations}

\label{Section - Computer Simulation}

In this section, the performance of the proposed approach is examined using
computer simulation studies. Table \ref{tb_sys_par}\ lists the system's
parameters. The RF carrier frequency is set to $1.3$ \textrm{GHz} for this
simulation. The DAC and ADC sampling interval $T_{s}$ is equal to $T_{c}$
and the unambiguous range $R_{u}$ is assumed to be $262$\ compressed range
bins while the \textrm{PRI} is $2R_{u}T_{c}$. The PN-code length/period $%
\mathcal{N}_{c}=15$ for pulse compression is also assumed.

\begin{table}
	\centering
	\caption{System Parameters}
	\label{tb_sys_par}
	\resizebox{\columnwidth}{!}{\begin{tabular}{|l|l|l|} 
			
			\hline
			\rowcolor{nb}  \textcolor{white}{Parameter}  & \textcolor{white}{Value}  & \textcolor{white}{Comment}                   \\
			\hline 
			\multirow{2}{*} {$F_c$ }                                          & \multirow{2}{*}{$1.3$ GHz }                 & \multirow{2}{*}{RF Carrier Frequency}                               \\
			& & \\
			
			\hline 
			\multirow{2}{*} {$T_c, T_s$ }                                        & \multirow{2}{*}{$T_s=T_c$ }                 & \multirow{2}{*}{Chip interval = Sampling frequency}                               \\
			& & \\						
			
			\hline 
			\multirow{2}{*} {$B$ }                                          & \multirow{2}{*}{\Large{$\frac{1}{Tc}$} }                 & \multirow{2}{*}{RF bandwidth}                               \\
			& & \\								
			
			\hline 
			\multirow{2}{*} {$\overline{N}, N $ }                           & \multirow{2}{*}{$5$ }                 & \multirow{2}{*}{Tx and Rx UCAs elements}                               \\
			& & \\

			\hline 
			\multirow{2}{*} {Rx SNR }                                       & \multirow{2}{*}{$20$ dB }                 & \multirow{2}{*}{Rx Signal to Noise Ratio}                               \\
			& & \\
			
			\hline 
			\multirow{2}{*} {SCR }                                       & \multirow{2}{*}{$-5$ dB }                 & \multirow{2}{*}{Signal to Clutter Ratio}                               \\
			& & \\

			\hline 
			\multirow{2}{*} {$\mathcal{N}_c$ }                              & \multirow{2}{*}{$15$ }                 & \multirow{2}{*}{PN-code length/period }                               \\
			& & \\
			
			\hline 
			\multirow{2}{*} {$T_p$ }                                        & \multirow{2}{*}{$T_{c}\mathcal{N}_{c}$ }                 & \multirow{2}{*}{Pulse duration }                               \\
			& & \\	
			
			\hline 
			\multirow{2}{*} {$R_u$ }                                        & \multirow{2}{*}{$262$}                 & \multirow{2}{*}{Unambiguous range \small{(in units of compressed range bins)} }                               \\
			& & \\		
			
			\hline 
			\multirow{2}{*} {\textrm{PRI}}                                         & \multirow{2}{*}{$2R_{u}T_c $ }                 & \multirow{2}{*}{Pulse Repetition Interval}                               \\
			& & \\			
			
			\hline 
			\multirow{2}{*} {\textrm{PRF}}                                          & \multirow{2}{*}{\Large{$\frac{1}{\textrm{PRI}}$} }                 & \multirow{2}{*}{Pulse Repetition Frequency}                               \\
			& & \\			
			
			\hline 
			\multirow{2}{*} {$L$}                                          & \multirow{2}{*}{\Large{$\frac{\textrm{PRI}}{Tc}$} }                 & \multirow{2}{*}{Compressed Range bins per PRI}                               \\
			& & \\	

			\hline 
			\multirow{2}{*} {$N_{s}$}                                          & \multirow{2}{*}{$256$ }                 & \multirow{2}{*}{Number of PRIs per CPI}                               \\
			& & \\				
			\hline
		\end{tabular}
	}
\end{table}%

\begin{figure}[h]
\centering
\includegraphics[width=0.9\columnwidth]{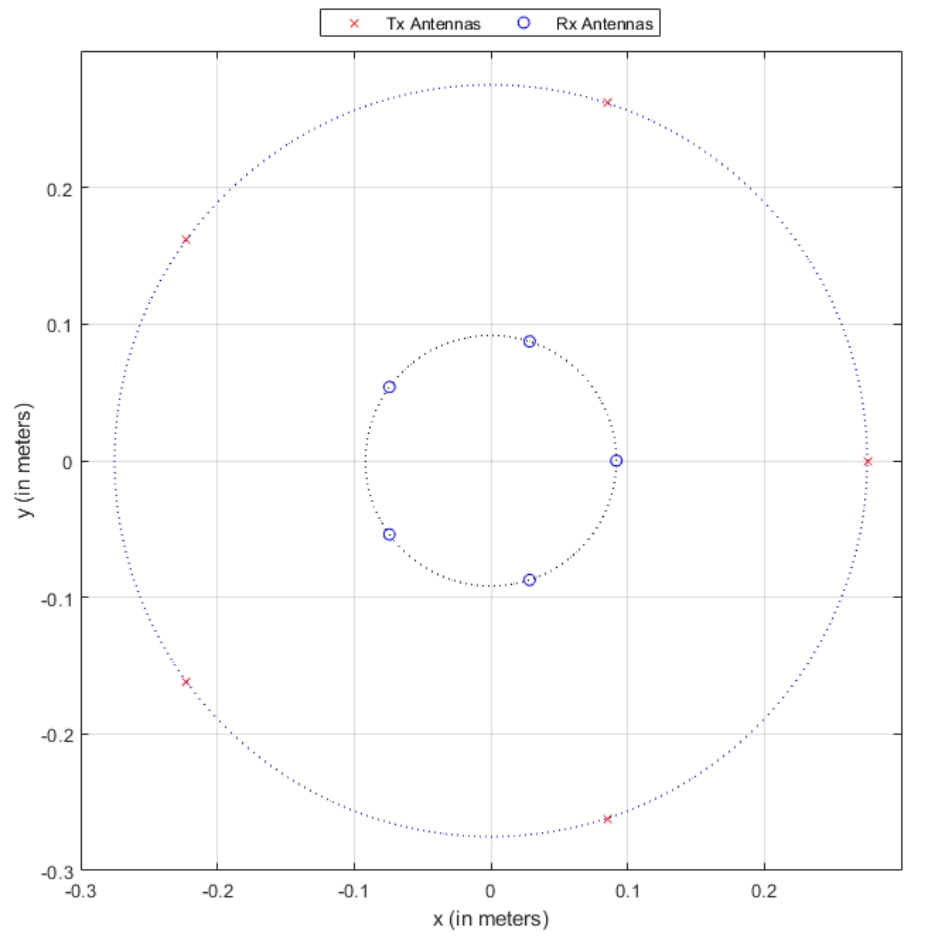}
\caption{Tx and Rx UCA}
\label{fig_tx_rx_uca}
\end{figure}

As shown in Figure \ref{fig_tx_rx_uca}, both the radar's Tx and Rx use
uniform circular arrays (UCA). The Rx antenna array elements are separated
by a distance $d=\frac{\lambda }{2}$ while the Tx elements are separated by $%
3d$ and each element has a unity gain ($G_{Tx}=G_{Rx}=1$). Equations \ref%
{eqn:tx_arr_g} and \ref{eqn:rx_arr_g} provide the Tx and Rx antenna array
geometries (Cartesian coordinates) on a 3-D real space in meters. The Tx and
Rx arrays or sites are separated by $L_{bi}=95$ units of compressed range
bins. All antenna array elements are along the x and y-axis and no element
is across the z-axis.

\begin{eqnarray}
\underline{\underline{\overline{\mathbf{r}}}} &=&%
\begin{bmatrix}
\underline{\overline{r}}_{1}, & \underline{\overline{r}}_{2}, & \ldots , & 
\underline{\overline{r}}_{\overline{N}}%
\end{bmatrix}
\notag \\
&=&%
\begin{bmatrix}
0.28, & 0.09, & -0.22, & -0.22, & 0.09 \\ 
0, & 0.26, & 0.16, & -0.16, & -0.26 \\ 
0, & 0, & 0, & 0, & 0%
\end{bmatrix}
\label{eqn:tx_arr_g}
\end{eqnarray}

\begin{eqnarray}
\underline{\underline{\mathbf{r}}} &=&%
\begin{bmatrix}
\underline{r}_{1}, & \underline{r}_{2}, & \ldots , & \underline{r}_{N}%
\end{bmatrix}
\notag \\
&=&%
\begin{bmatrix}
0.092, & 0.028, & -0.074, & -0.074, & 0.028 \\ 
0, & 0.087, & 0.054, & -0.054, & -0.087 \\ 
0, & 0, & 0, & 0, & 0%
\end{bmatrix}
\notag \\
&&  \label{eqn:rx_arr_g}
\end{eqnarray}

\noindent where $d$ is defined as: 
\begin{equation*}
d=\frac{\lambda }{2}=\frac{c}{2F_{c}}
\end{equation*}

\begin{figure}[h]
\centering
\includegraphics [width=0.85\columnwidth]{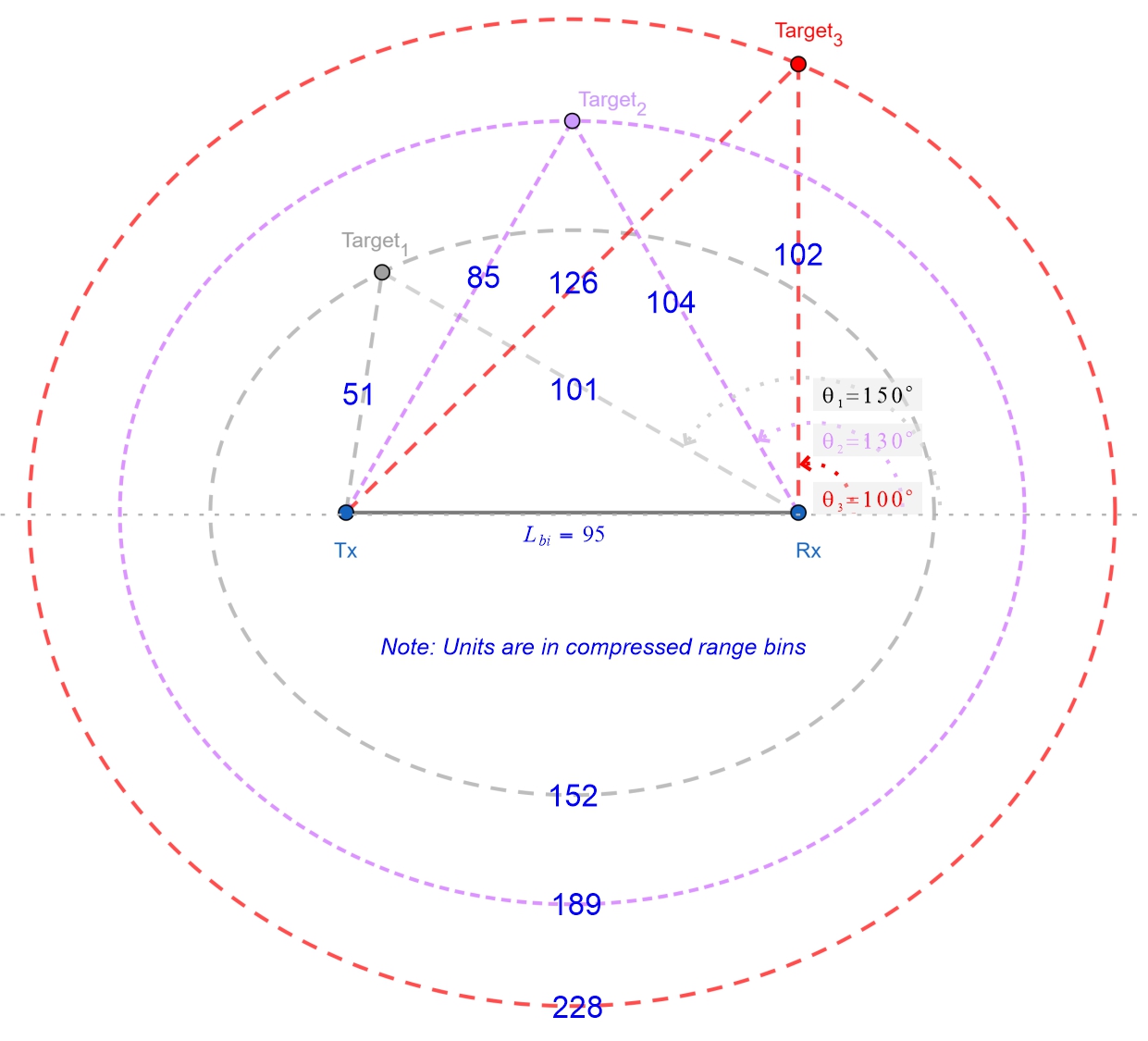}
\caption{Targets' Geometry}
\label{fig_tar_geo}
\end{figure}

Figure \ref{fig_tx_rx_uca} shows both antenna arrays with reference to their
own local reference points. The figure does not show the distance between
the Tx and Rx array. Figure \ref{fig_tar_geo} shows a bistatic radar system
with Tx and Rx Cartesian coordinate location and three targets on the (x,y)
plane. The elevation angle is assumed zero.Table \ref{tb_tar_par} summarises
the geometric parameters of three targets along with their RCS model and
average values. From this table, RCS values in conjunction with the Swerling
probability density functions, are used to generate the received reflections
(random numbers) for each target.\\
As described in Section 2, the transmitted waveform with PN-codes (either
m-code or Gold codes) for pulse compression provides orthogonality within a
PRI period. The period could be adjusted for required target illumination
and the resulting signal-to-noise ratio.
\begin{table*}[ht!]
\caption{Targets Parameters}
\label{tb_tar_par}\centering%
\begin{tabular}{||l||l||l||l||l||}
\hline
\rowcolor{nb} \textcolor{white}{Parameter} & \textcolor{white}{Target-1} & %
\textcolor{white}{Target-2} & \textcolor{white}{Target-3} & %
\textcolor{white}{Comments} \\ \hline
$R_{Tx}$ & $51$ & $85$ & $126$ & Tx to Target range (in compressed range
bins) \\ \hline
$R_{Rx}$ & $101$ & $104$ & $102$ & Rx to Target range (in compressed range
bins) \\ \hline
$R_{bi}$ & $152$ & $189$ & $228$ & Bistatic range = $R_{Tx}+R_{Rx}$ (in
compressed range bins) \\ \hline
$\theta $ & $150^{\circ}$ & $130^{\circ}$ & $100^{\circ}$ & Direction of
Arrival (DOA) \\ \hline
&  &  &  &  \\[-1.15em] 
$\overline{\theta }$ & $81.20^{\circ}$ & $70.83^{\circ}$ & $52.31^{\circ}$ & 
Direction of Departure(DOD) \\ \hline
$\zeta $ & $68.80^{\circ}$ & $59.17^{\circ}$ & $47.69^{\circ}$ & Bistatic
Angle \\ \hline
RCS & $1.0$ m$^{2}$ & $1.5$ m$^{2}$ & $2$ m$^{2}$ & Radar Cross-Section \\ 
\hline
Swerling Model & $1^{\text{st}}$ model & $2^{\text{nd}}$ model & $3^{\text{rd%
}}$ model & Swerling model for RCS \\ \hline
$v$ & $-60$ m/sec & $20$ m/sec & $60$ m/sec & Radial Velocity (bistatic) \\ 
\hline
\end{tabular}%
\end{table*}
\FloatBarrier

The radar also is assumed to operate in the presence of ground clutter and
AWGN. The ground clutter can be modelled using various other PDFs such as
K-distribution and Weibull etc. It is often challenging to model and filter
various types of clutters accurately because different objects/terrains
cannot be represented by a single model. One effective approach is to
transform the clutter PDFs to Gaussian i.e. transform the coloured clutter
to white noise (isotropic) and then filter it out the same way as white
noise \cite{Ren2019}. The clutter filtering for this experiment is
implemented as described in \cite{Ren2019}.

\begin{figure*}[tbp]
\centering
\includegraphics[width=0.82\textwidth]{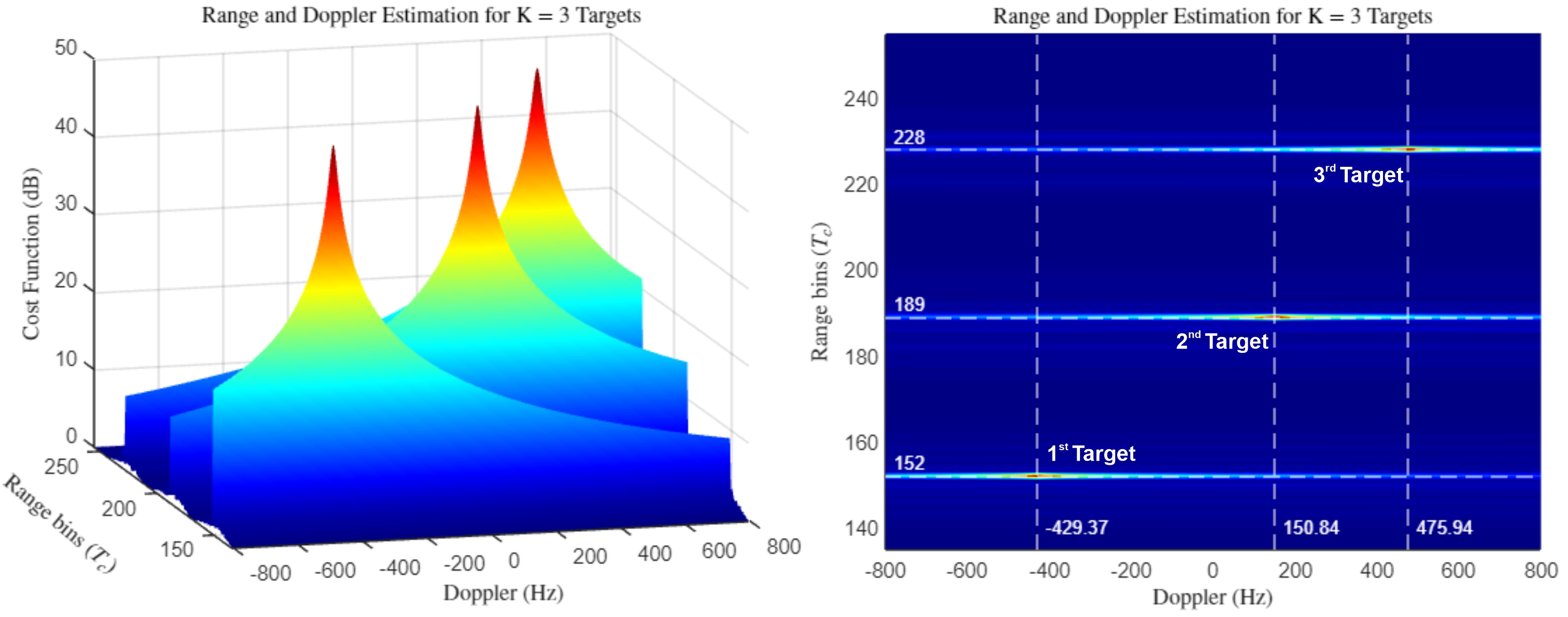}
\caption{Range and Doppler Estimation.}
\label{fig_rng_dop_est}
\end{figure*}

\begin{figure*}[tbp]
\centering
\includegraphics[width=0.82\textwidth]{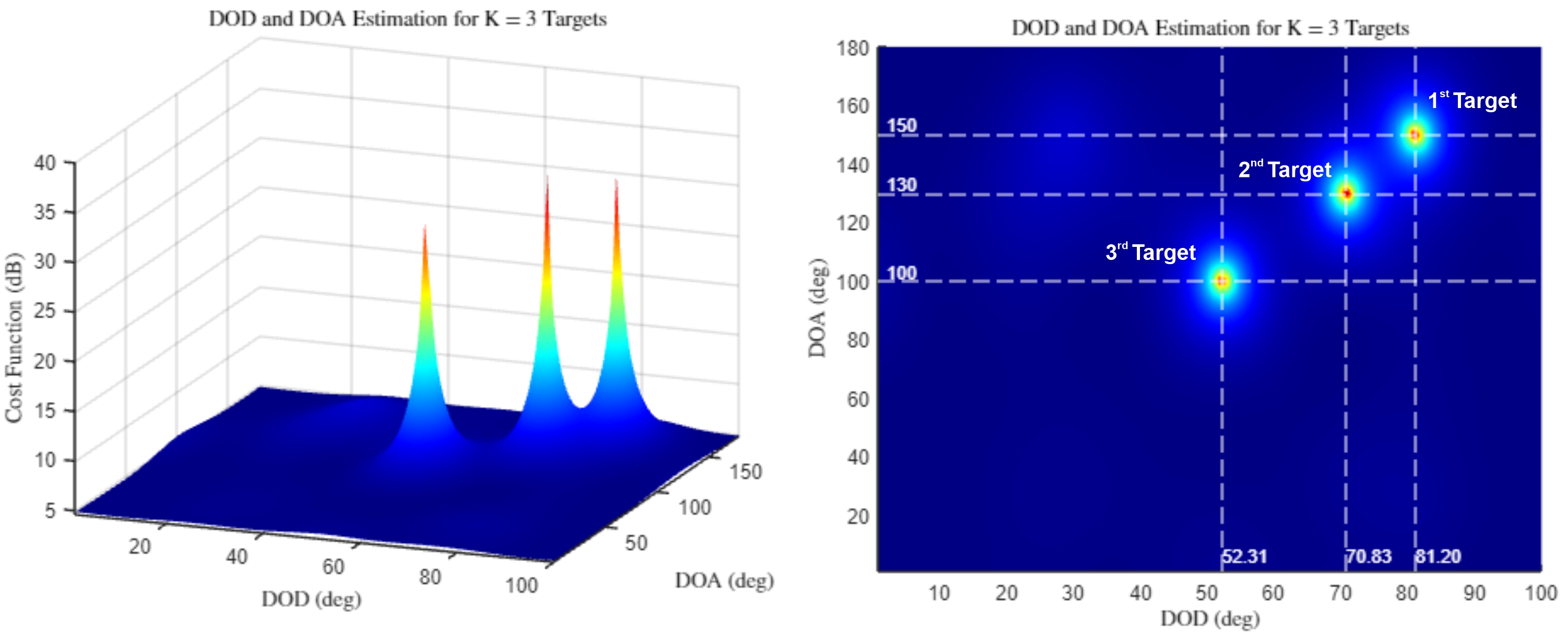}
\caption{DOA and DOD Estimation.}
\label{fig_al1_est_doa_dod}
\end{figure*}

\subsection{Range and Doppler Estimation}

The range and Doppler can be estimated jointly by just utilising the
temporal dimension of the data and array manifold. Therefore, the cost
function of Equation \ref{eq_rd_02} is used for the estimation. Figure \ref%
{fig_rng_dop_est} shows that the peaks are sharp and the estimates are very
close to the range and Doppler resolution error. The results are summarised
in Table \ref{tb_birng_dop_est} below.

\begin{table}[h!]
\caption{Bistatic Range and Doppler Estimations}
\label{tb_birng_dop_est}\centering%
\begin{tabular}{||l||l||l||l||l||}
\hline
\rowcolor{nb} & \multicolumn{2}{c||}{\cellcolor{nb}\textcolor{white}{$R_{bi}
(T_c)$}} & \multicolumn{2}{c||}{\cellcolor{nb}\textcolor{white}{Doppler (Hz)}
} \\ \hline\hline
\rowcolor{nb} \textcolor{white}{Target} & \textcolor{white}{True} & %
\textcolor{white}{Estimated} & \textcolor{white}{True} & %
\textcolor{white}{Estimated} \\ \hline
$1$ & $152$ & $152$ & $-429.37$ & $-428.37$ \\ \hline
$2$ & $189$ & $189$ & $150.84$ & $151.13$ \\ \hline
$3$ & $228$ & $228$ & $475.94$ & $475.13$ \\ \hline
\end{tabular}%
\end{table}

\subsection{DOA and DOD Estimation}

A joint DOA and DOD estimation is performed using the cost function of
Equation \ref{cf_doa_dod_02}. As shown in Figure \ref{fig_al1_est_doa_dod}
and Table \ref{tb_al1_doa_dod_est}\ below, the estimates are very close.

\begin{table}[h!]
\caption{DOA and DOD Estimation}
\label{tb_al1_doa_dod_est}\centering%
\begin{tabular}{||l||l||l||l||l||}
\hline\hline
\cellcolor{nb}\vspace{-1em} & \multicolumn{2}{||l||}{\cellcolor{nb}} & 
\multicolumn{2}{||l||}{\cellcolor{nb}} \\ 
\cellcolor{nb} & \multicolumn{2}{||c||}{\cellcolor{nb}\textcolor{white}{$%
\overline{\theta}$}} & \multicolumn{2}{||c||}{\cellcolor{nb}%
\textcolor{white}{$\theta$}} \\ \hline\hline
\cellcolor{nb}\textcolor{white}{Target} & \cellcolor{nb}%
\textcolor{white}{True} & \cellcolor{nb}\textcolor{white}{Estimated} & %
\cellcolor{nb}\textcolor{white}{True} & \cellcolor{nb}%
\textcolor{white}{Estimated} \\ \hline\hline
$1$ & $81.20^{\circ }$ & $81.20^{\circ }$ & $150.00^{\circ }$ & $%
150.01^{\circ }$ \\ \hline\hline
$2$ & $70.83^{\circ }$ & $70.83^{\circ }$ & $130.00^{\circ }$ & $%
130.01^{\circ }$ \\ \hline\hline
$3$ & $52.31^{\circ }$ & $52.31^{\circ }$ & $100.00^{\circ }$ & $%
100.02^{\circ }$ \\ \hline\hline
\end{tabular}%
\end{table}

\subsection{RMSE and Comparison}

The algorithm, as denoted by Equation \ref{cf_doa_dod_02}, has been
rigorously assessed utilizing Monte Carlo simulations and the Root Mean
Square Error (RMSE) method. A comparative analysis has been conducted
against the standard bistatic radar estimation method, given in Appendix A.
The error assessment for the parameter of interest $p$ is defined as per the
following equation:%
\begin{equation}
RMSE_{p}=\frac{1}{K}\overset{K}{\underset{k=1}{\sum }}\sqrt{\mathcal{E}%
\{\left\vert \widehat{p}_{k}-p_{k}\right\vert ^{2}\}}
\end{equation}%
Where $p_{k}$ represents the general parameter of interest for the $k$-th
target.

\begin{figure}[h]
\centering
\includegraphics[width=0.85\columnwidth]{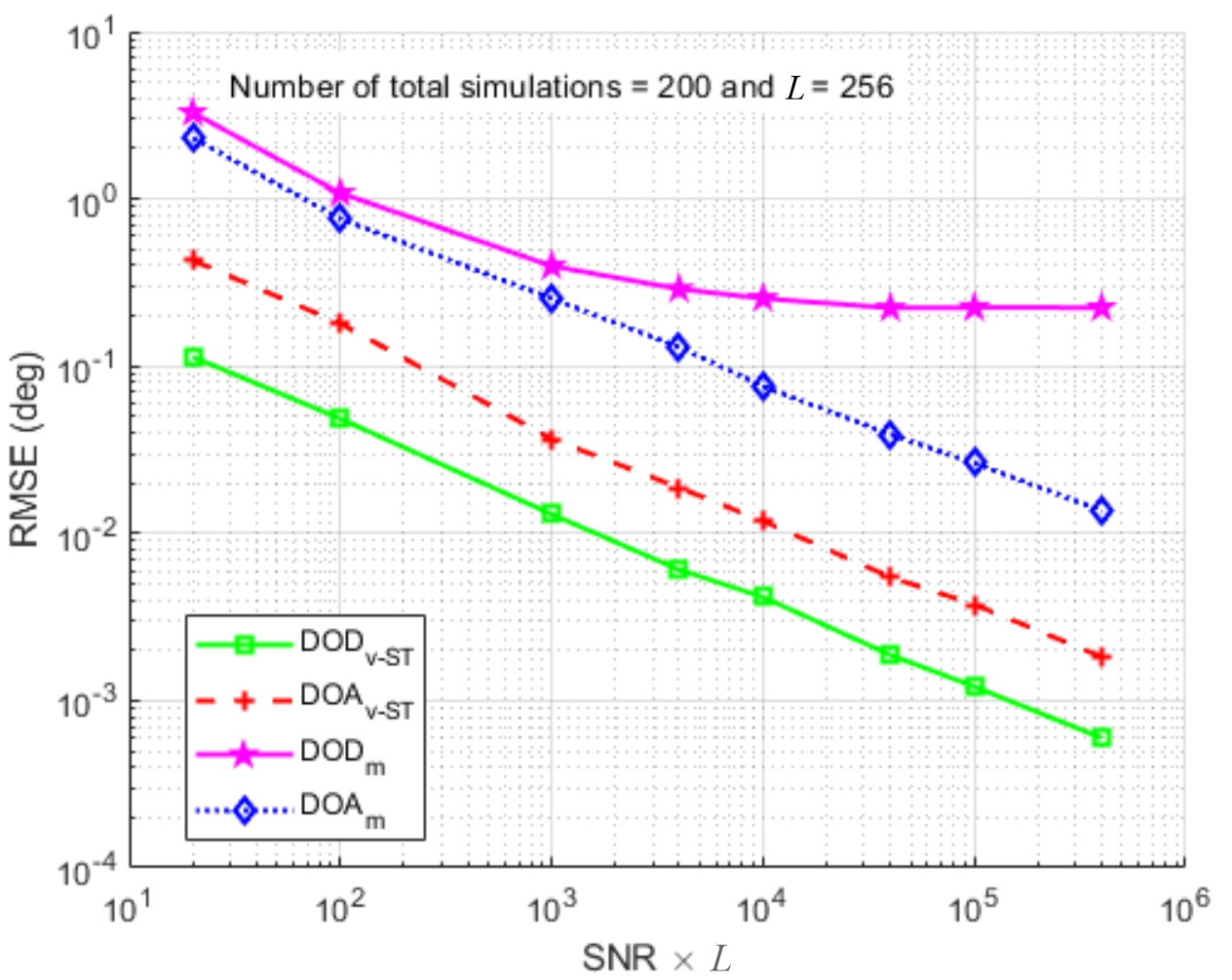}
\caption{DOA and DOD RMSE}
\label{fig_09_rmse}
\end{figure}

The comparison depicted in Figure \ref{fig_09_rmse} is annotated with
subscript letters corresponding to the DOD and DOA values in the legend,
represented as follows:

\begin{itemize}
\item DOD$_{\mathrm{v}\text{-ST}}$: DOD with v-ST algorithm.

\item DOA$_{\mathrm{v}\text{-ST}}$: DOA with v-ST algorithm.

\item DOD$_{\mathrm{m}}$: DOD with {\small MUSIC algorithm and bistatic
geometry.}

\item DOA$_{\mathrm{m}}$: DOA with MUSIC algorithm.
\end{itemize}

The performance depicted in the figure demonstrates the superior efficacy of
the proposed virtual Spatiotemporal (v-ST) algorithm when compared to
conventional bistatic estimation techniques. It is noteworthy that the DOD$_{%
\mathrm{m}}$ exhibits a floor-level trend beyond the mid-trace, attributable
to range floor-level quantization error limitations.

\section*{Conclusions}

In conclusion, this paper investigated the application of novel subspace
algorithms for parameter estimation in MIMO bistatic pulse radar systems.
The proposed approach uses a unique PN-code for each transmit chain,
maximising the observation space by utilizing both the Tx and Rx observation
spaces. The channel model incorporates $K$ moving targets with
Swerling-distributed RCS and accounts for clutter and noise. The received
signal model is constructed, and a 3D radar data cube is assembled. This
data undergoes processing through Spatiotemporal and Virtual Spatiotemporal
manifold extenders to increase the degree of freedom and improved parameter
estimation.

Computer simulations were conducted to evaluate the performance of the
proposed direction-of-arrival (DOA) and direction-of-departure (DOD)
estimation algorithm. Using Monte Carlo simulations and Root Mean Square
Error (RMSE) as a metric, a comparison with the bistatic MIMO radar
equations in conjunction with the MUSIC algorithm revealed the superior
performance of the proposed system.

\section*{Acknowledgments}

This research would not have been possible without the funding from Saab
Research Sweden/UK. The authors, in particular, would also like to express
their deepest gratitude to Prof. Anders Silander (Saab) and Malin Svahn
(Saab, VP Director of Innovation Programmes) for their continuous support
and encouragement.

\bibliographystyle{IEEEtran}
\bibliography{library_FTC_paper_01}

\appendix{}

\subsection{Bistatic Estimation Process}

\begin{enumerate}
\item Compute the bistatic range estimation ({\normalsize $R_{bi}$}) using
Equation \ref{eq_rd_02}.

\item Using $L_{bi}$\ and {\normalsize $R_{bi}$, }determine the following
bistatic target's ellipse parameters:

\begin{itemize}
\item semi-major axis $a$

\item semi-minor axis $b,$ and

\item its eccentricity $\varepsilon $,

using Equations \ref{eq_ell_ak}-\ref{eq_ell_ek} of Appendix B.
\end{itemize}

\item Calculate the Tx to Target Range ($R_{Tx}$) and Target to Rx Range ($%
R_{Rx}$) using Equations \ref{eq_fur_range_Rtx} and \ref{eq_fur_range_Rrx}
of Appendix B.

\item Using the MUSIC algorithm estimate the Direction of Arrival (%
{\normalsize DOA$_{\mathrm{m}}$}).

\item Estimate the Direction of Departure (DOD$_{\mathrm{m}}$) using
Equation \ref{eq_dod_mg}\ given in Appendix B
\end{enumerate}

\subsection{Bistatic Equations}

The estimated target delay parameters $d_{k}$ can be converted to bistatic
range $R_{bi}$\ as follows.\ 
\begin{eqnarray}
R_{bi,k} &=&d.c.T_{c}\mid _{d=d_{k}},\forall k  \label{eq_fur_range_Rbi} \\
R_{bi,k} &=&R_{Tx,k}+R_{Rx,k}
\end{eqnarray}%
With reference to Figure \ref{fig_ellipse_01} which shows a bistatic ellipse
and its parameters, the values $R_{bi,k}$ and $L_{bi}$ can be used to
estimate the parameters $a_{k}$, $b_{k}$ and $\varepsilon _{k}$ as follows:

\begin{figure}[h]
\centering
\includegraphics[width=0.82\columnwidth]
{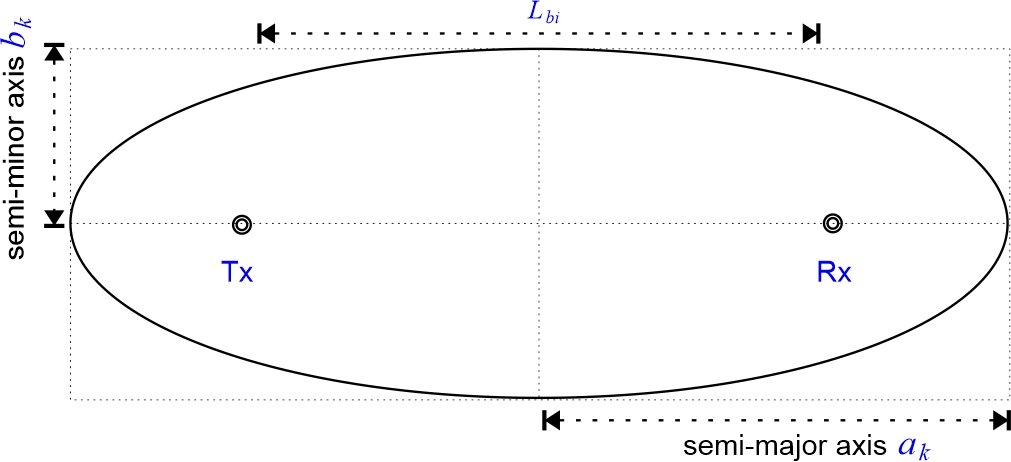}
\caption{Ellipse notations}
\label{fig_ellipse_01}
\end{figure}

\begin{eqnarray}
\text{semi-major axis }a_{k} &=&\frac{R_{bi,k}}{2},\forall k
\label{eq_ell_ak} \\
\text{semi-minor axis }b_{k} &=&\sqrt{R_{bi,k}^{2}-\frac{L_{bi}^{2}}{4}}
\label{eq_ell_bk} \\
\text{eccentricity ratio }\varepsilon _{k} &=&\frac{L_{bi}/2}{a_{k}}
\label{eq_ell_ek}
\end{eqnarray}%
Furthermore, the bistatic ranges{\normalsize \ $R_{Rx}$\ (Target to Rx) and $%
R_{Tx}$ (Tx to Target) for each target can be estimated using the following
equations:%
\begin{eqnarray}
R_{Rx,k}\ &=&\frac{a_{k}(\varepsilon _{k}^{2}-1)}{\varepsilon _{k}\cos
(\theta _{k})+1},\forall k  \label{eq_fur_range_Rtx} \\
R_{Tx,k} &=&R_{bi,k}-R_{Rx,k}  \label{eq_fur_range_Rrx}
\end{eqnarray}%
Finally, using the above equation, the DOD }$\overline{\theta }_{k}$%
{\normalsize \ $\forall k$ can be estimated as follow:%
\begin{equation}
\overline{\theta }_{k}=\arccos \left( \frac{1}{\varepsilon _{k}}\left( \frac{%
a_{k}(\varepsilon _{k}^{2}-1)}{R_{Tx,k}}+1\right) \right)  \label{eq_dod_mg}
\end{equation}%
}
\vspace{-2em}
\begin{IEEEbiography}
[{\includegraphics[width=1.1in,height=1.25in,clip,keepaspectratio]{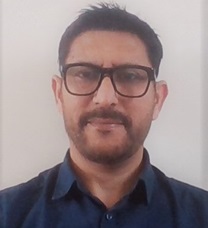}}]
{Nadeem Dar} received his BEng (Honors) degree in Communication Systems Engineering from the University of Westminster, London,  in 2000 and his MSc degree in Electronic and Electrical Engineering from Anglia Ruskin University, Cambridge, in 2019. Currently, he is pursuing his PhD degree within the Communications and Signal Processing Group at Imperial College London's Department of Electrical and Electronic Engineering. His research focuses on advanced topics in bistatic/multistatic arrayed MIMO radar systems, with a particular emphasis on array signal processing, super-resolution parameter estimation, and space-time array processing.
\end{IEEEbiography}

\begin{flushleft}
\begin{IEEEbiography}
[{\includegraphics[width=1.1in,height=1.25in,clip,keepaspectratio]{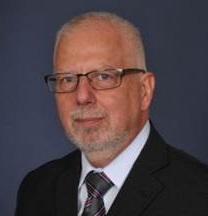}}]
{Athananssios Manikas,}please see: \newline https://skynet.ee.ic.ac.uk/manikas.html
{\color{white} .......  ........ ....... ....... ...... ........ ........ ..... ...... ..... ........ ...... ...... .... ........... ........

} 
\end{IEEEbiography}
\end{flushleft}

\end{document}